\documentclass[]{article}
\usepackage{cite}
\usepackage{graphicx}
\usepackage{amsfonts,amssymb,amsmath}

\usepackage[caption=false,font=footnotesize]{subfig}
\usepackage[export]{adjustbox}

\usepackage[utf8]{inputenc} 
\usepackage[T1]{fontenc}    
\usepackage{hyperref}       
\usepackage{url}            
\usepackage{booktabs}       
\usepackage{amsfonts}       
\usepackage{nicefrac}       
\usepackage{microtype}      

\date{}

\usepackage[left=2.30cm, right=2.30cm, top=2.30cm, bottom=2.30cm]{geometry}

\title{Intelligent Reflecting Surfaces with Spatial Modulation:\\An Electromagnetic Perspective}

\author{
	Okan Yurduseven, \textit{Senior Member, IEEE}, \\
	Stylianos D. Assimonis, \\
	and Michail Matthaiou, \textit{Senior Member, IEEE}
	\thanks{The authors are with the Institute of Electronics, Communications and Information Technology (ECIT), Queen`s University Belfast, Belfast, BT3 9DT, U.K., (e-mail: \{okan.yurduseven, s.assimonis, m.matthaiou\}@qub.ac.uk).}
}

\begin{document}

\maketitle

\begin{abstract}Electromagnetic wave control using the concept of a reflecting surface is first studied as a near-field and a far-field problem. Using a secondary source present in a wireless communication environment, such as a backscatter tag, it is possible to leverage the incoming radiation from the source as a reference-wave to synthesize the desired wavefront across the reflecting surface, radiating a field of interest. In this geometry, the phase grating, which is synthesized using an array of sub-wavelength unit cells, is calculated by interacting the incident reference-wave and the desired wavefront, similar to a hologram. When illuminated by the reference-wave, the reflected wavefront from the calculated phase grating is guaranteed to constructively add in the direction of the desired radiation (beam-steering in the far-field) and also focus at the intended depth (beam-focusing in the radiative near-field). Leveraging a dynamic modulation mechanism in the context of an intelligent reflective surface (IRS) illuminated by a backscatter tag, later, we present that one can selectively focus and defocus at an arbitrarily positioned receiver within the 3D field of view of the reflecting surface. This enables the control of the amplitude of the radiated electric field at the receiver location, paving the way for a spatial modulation mechanism by means of reconfiguring the reflecting surface in a backscattered wireless communication environment. In addition to the phase modification approach on a unit cell level to reconfigure the aperture radiated wavefronts, we finally present a time varying IRS concept making use of a time-delay based approach relying on a delay adjustment between the reflection coefficients of the IRS' unit cell lattices. 
\end{abstract}




\section{Introduction}

{I}{ntelligent} reflecting surfaces (IRS) have recently emerged as a promising candidate for offering ubiquitous connectivity and seamless coverage in beyond 5G and 6G networks \cite{Liaskos, Dai, Wu}. IRS can be regarded as a physical evolution of massive MIMO, where hundreds (or even thousands) of antenna modules (or meta-atoms) are coated on the walls of buildings, factories or even aircrafts \cite{Zhang}. 

IRS offers, in theory, some key advantages over concurrent wireless technologies, such as relaying and cell-free massive MIMO. For example, they do not suffer from the fundamental problem of \textit{self-interference}, induced by full-duplex relaying; they also do not impose additional backhaul requirements as cell-free massive MIMO does \cite{Ngo}. Most importantly, their operational power consumption can be rather small, particularly when combined with passive elements \cite{Zhang, Liaskos2}. For these reasons, it comes as no surprise that the related literature has been fast expanding with numerous papers appearing over the last year. 
In this context, it is worth mentioning the stream of publications on channel estimation for IRS \cite{channelest1,channelest2}, precoding design \cite{design1,design2} and physical prototyping \cite{Dai,Jin}.    

Nevertheless, the vast majority of papers investigate the IRS-related problems using standard information theoretic tools and techniques by ignoring inherent electromagnetic phenomena. Yet, it is an indisputable fact that a holistic performance analysis requires amalgamation of information theoretic and electromagnetic tools. Testament to this, \cite{Emil2} recently showed that the signal-to-noise ratio (SNR) of IRS grows quadratically with the number of array elements only in the far-field and also that that the IRS setup cannot achieve a higher SNR than the corresponding massive MIMO setup, if the array sizes are equal. 

In this paper, we are harnessing knowledge of the electromagnetic characteristics to develop a new spatial communication protocol suitable for IRS. As was pointed out in the overview article \cite{diRenzo}, the development of communication protocols is one of key challenges pertaining to the successful roll-out of IRS. Our analysis is general enough as it is applicable to both near-field and far-field scenarios. Using a set of closed-form analytical equations, we present the design process of reflecting surfaces and verify their operation in both near-field and far-field regions using a full-wave electromagnetic solver, CST Microwave Studio.  The ability to control  the radiation magnitude in depth (beam-focusing) as well as the direction of the radiated wavefront (beam-steering), results in a full 3D beam control capability for IRS type apertures. This can be particularly useful in backscatter type wireless communication environment with the reflecting surface achieving a selective focusing over a receiver unit placed at an arbitrary position. Leveraging the dynamically reconfigurable, selective beam-focusing concept, we present a spatial modulation mechanism. As well as adopting a phase modulation approach to reconfigure the reflection phases of the unit cells forming the IRS, we also present a time-delay approach to reconfigure the radiation wavefront. Leveraging the time-delay based approach, the implementation of the spatial modulation scheme can be easily performed by appropriate reflection phase time-delay adjustment of the reflection coefficient across each IRS’ lattice of unit cells.  

The specific contributions of the paper are as follows:
\begin{enumerate}
    \item We first develop a novel mechanism for 3D  beam forming in the near and far-field of an IRS. A particularly important aspect of the presented study is the dynamic beam-focusing concept, which goes beyond the conventional beam-steering capability of IRS architectures \cite{bjornson2019intelligent,Wu,8723525,9087848,8742603}.
    \item As a next step, we present the design of a new spatial modulation scheme for a near-field focusing IRS. In this scheme, focusing can be achieved by choosing
the correct reflection phases for the unit cells, whilst the defocusing of the radiated field can be achieved by randomizing the unit cell phase responses across the reflecting surface aperture.
    \item Using basic Fourier theory, we finally show that a planar \textit{time-varying} IRS can steer the reflected power to a specific direction by simple time delay adjustment across the reflection coefficients. This very simple topology is perfectly suitable for low-cost, low-power backscattering as the unit cells of the IRS are designed once with their reflection coefficients, which periodically alternate between -1 and 1, whilst a microcontroller can cater for beamforming.
\end{enumerate}

The outline of the paper is as follows: In Section II, we present the design principles of reflecting surface type apertures from an electromagnetic perspective. Particularly, we focus on two different antenna regions, far-field and near-field (radiative) from a design perspective, and present two different electromagnetic beam forming operations in these respective fields; namely beam-steering in the far-field and beam-focusing in the near-field. 
 Building on the design process laid out in Section II, in Section III, we present the concept of dynamically modifying the phase responses of the unit cells to achieve selective beam-focusing for a wireless communication environment. In Section IV, we develop a novel technique to realize low-complexity backscattering through IRS by means of a time-delay based approach. Finally, in Section V, we present the concluding remarks.    

\section{Intelligent Reflecting Surfaces for Electromagnetic Wave Control}
\subsection{Far-Field Beam-Steering}
A reflecting surface can be considered an example of a metasurface hologram synthesized using an array of sub-wavelength unit cells reflecting the incoming electromagnetic radiation produced by a secondary source \cite{7792160,silva2019metasurface,8613936}. Using a reflecting surface, one can synthesize a desired aperture wavefront producing a radiation pattern of interest in the far-field. Beam synthesis from a reflecting surface relies on a holographic principle in which a secondary source illuminating the reflecting surface acts as a reference-wave. Upon reflection from the surface, the reflected wavefront is modulated by the phase profile of the reflecting surface into an objective function, which is the projection of the desired far-field radiation pattern on the aperture of the reflecting surface. As a secondary source launching the reference-wave, one can use a simple patch antenna, which has been widely adopted in the literature as a backscatter tag \cite{lu2018microwave,6646592,7481962}. Treating the patch antenna as a collection of point source elements representing a surface current distribution, $J(r')$, the electric field (or E-field) incident on the reflecting surface aperture can be calculated as follows:

\begin{equation}
    E(r)=\int_{r'}J(r')\frac{e^{-jk_0|r-r'|}}{4\pi|r-r'|}dr'.
    \label{Equation_1}
\end{equation}

In (\ref{Equation_1}), $J(r')$ denotes the surface current distribution on the secondary antenna (patch tag) illuminating the reflecting surface, $r'(x',y',z')$ denotes the coordinates of the secondary source, and $r(x,y,z)$ denotes the coordinates of the reflecting surface. From the holographic definition, the far-field Array Factor (AF) for the reflecting surface can be given as follows: 

\begin{equation}
    AF(\theta,\phi)=e^{-jk_0[x_i\sin(\theta)\cos(\phi)+y_i\sin(\theta)\sin(\phi)]}.
    \label{Equation_2}
\end{equation}

\begin{figure}[t!]
	\centering
	\includegraphics[width=7.5 cm,trim=0cm 0cm 0cm 0cm]{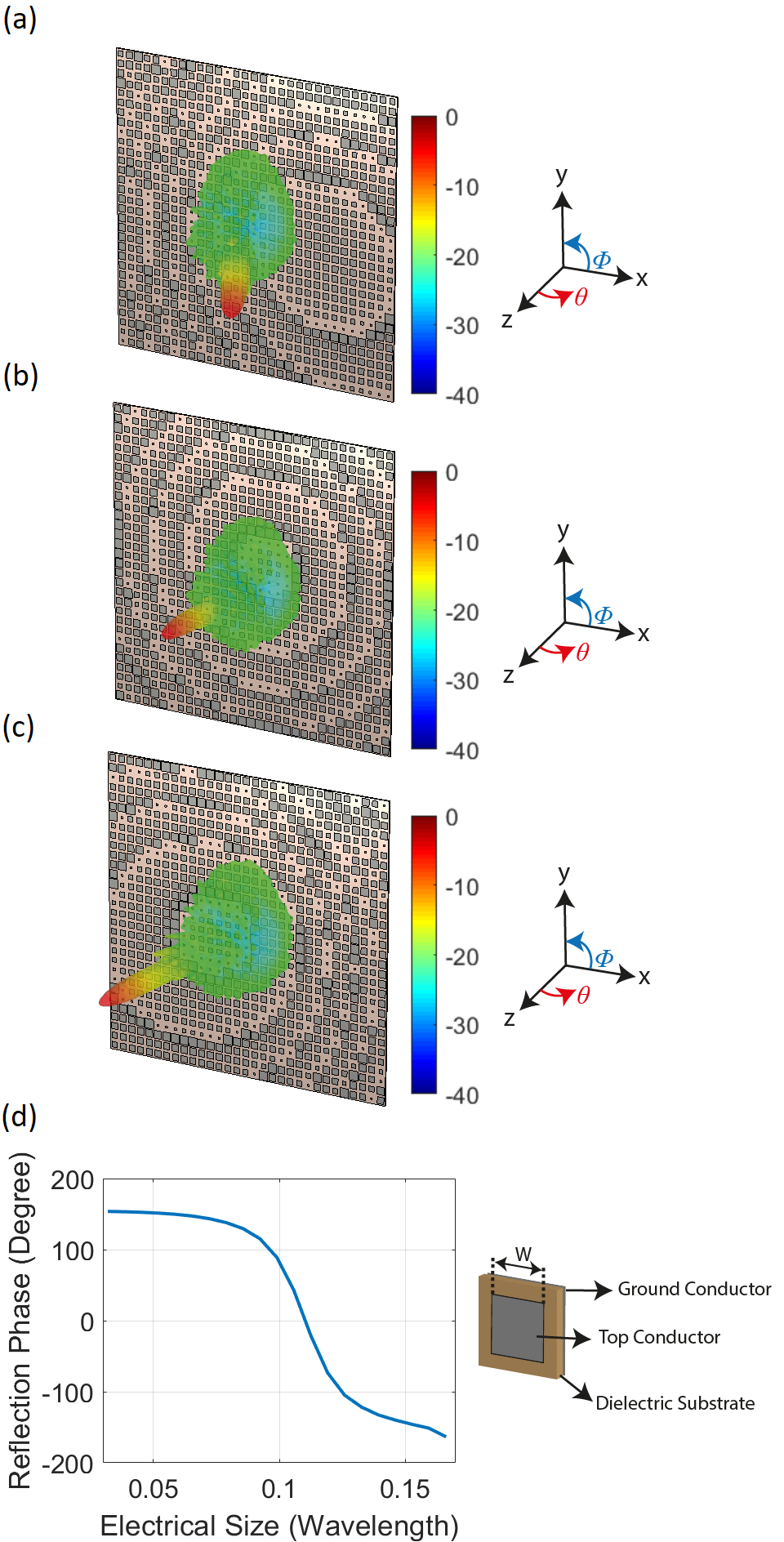}
	\caption{Beam-steering using a reflecting surface (a) ($\theta_1=-30^\circ$,$\phi_1=-30^\circ$) (b) ($\theta_2=0^\circ$,$\phi_2=0^\circ$) (c) ($\theta_3=30^\circ$,$\phi_3=30^\circ$) (d) unit cell reflection phase diagram as a function of varying electrical size ($W$) expressed in free-space wavelength. The dielectric substrate material is Rogers 4003 with a dielectric constant $\epsilon_r=3.38$. Colorbar is in dB scale.}
	\label{Figure_1}
\end{figure}

In (2), $k_0$ is the wavenumber in free-space, 
whilst $x_i$ and $y_i$ denote the position of the $i^{th}$ unit cell along the $x$-axis and $y$-axis, respectively. The steering angle $\theta$ is defined with respect to the broadside direction of the reflecting surface (z-axis) with $\theta=0^\circ$ representing the $z$-axis. The angle $\phi$ is defined in the transverse plane ($xy$-plane) with $\phi=0^\circ$ remaining along the $x$-axis. For a reflecting surface with $N$ x $N$ elements, $i=1\rightarrow{N^2}$. Hence, the total phase profile on the aperture of the reflecting surface is (assuming that the aperture is at $z=0$):

\begin{equation}
    \xi(x_i,y_i)=\angle{E(x_i,y_i)e^{-jk_0[x_i\sin(\theta)\cos(\phi)+y_i\sin(\theta)\sin(\phi)]}}.
    \label{Equation_3}
\end{equation}

Analyzing (\ref{Equation_2}) and (\ref{Equation_3}), it is evident that, a constructive interference can be achieved in a given ($\theta,\phi$) direction by maximizing the exponential term (and hence the AF) for the steering angle of interest. To achieve this, the phase advance introduced by the unit cells of the reflecting surface must be chosen as the complex conjugate of (\ref{Equation_3}) as follows: 

\begin{equation}
    \xi_{\textrm{unit}}(x_i,y_i)=\xi^*(x_i,y_i).
    \label{Equation_4}
\end{equation}

\begin{figure}[t!]
	\centering
	\includegraphics[width=6.5 cm,trim=0cm 0cm 0cm 0cm]{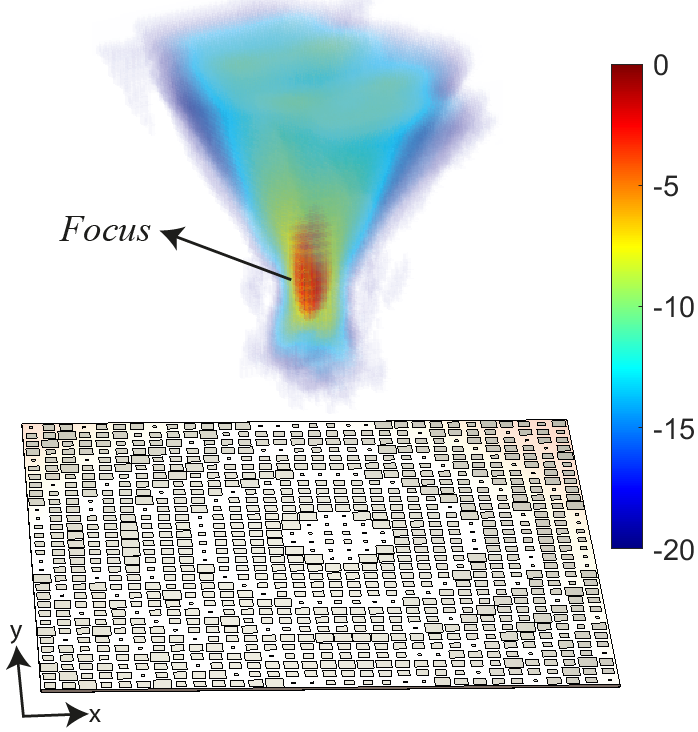}
	\caption{Beam-focusing using a reflecting surface. The E-field is focused in the broadside direction ($\theta=0^\circ, \phi=0^\circ$) at $d=0.45$m distance remaining in the radiative near-field region of the reflecting surface. Colorbar is in dB scale.}
	\label{Figure_2}
\end{figure}

\begin{figure}[t!]
	\centering
	\includegraphics[width=7 cm,trim=0cm 0cm 0cm 0cm]{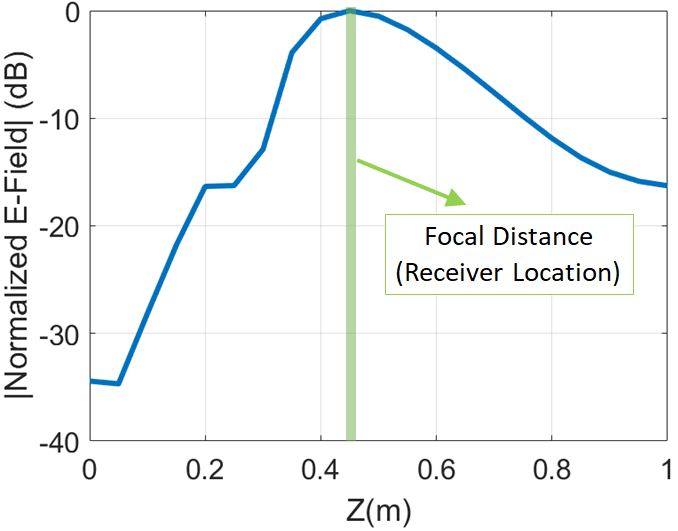}
	\caption{Cross-section of the E-field magnitude along the depth axis ($z$-axis) of the reflecting surface.}
	\label{Figure_3}
\end{figure}

In (\ref{Equation_4}), the symbol $^*$ denotes the complex conjugate operator. This selection ensures that when illuminated by the reference wave of (\ref{Equation_1}), the reflected wavefront from the unit cells of the reflecting surface will exhibit a constructive interference in the direction of the intended beam-steering angle, ($\theta$, $\phi$). To validate this scenario, in Fig. \ref{Figure_1}, we design several reflecting surface apertures to steer the beam in arbitrarily selected directions, ($\theta_1=30^\circ$, $\phi_1=30^\circ$), ($\theta_2=0^\circ$, $\phi_2=0^\circ$) and ($\theta_3=-30^\circ$, $\phi_3=-30^\circ$), respectively. The reflecting surface geometries together with the radiated far-field patterns are shown in Figs. \ref{Figure_1}a-c while the unit cell topology with the reflection phase diagram is shown in Fig. \ref{Figure_1}d.

The reflecting surface aperture in Fig. \ref{Figure_1} is illuminated by a patch tag radiating an E-field (reference-wave) incident on the reflecting surface at an arbitrarily selected angle of ($\theta_{\textrm{inc}}=10^\circ, \phi_\textrm{inc}=10^\circ$). As can be seen in Fig. \ref{Figure_1}, interacting the incident reference-wave with the calculated unit cells exhibiting the reflection phase profile of (\ref{Equation_4}) realized using the unit cell topology shown in Fig. \ref{Figure_1}(d), the radiated wavefronts by the reflecting surface aperture exhibit highly directive far-field beam patterns, pointing in the desired beam-steering directions.  

\subsection{Near-Field Beam-Focusing}
In addition to beam-steering, which is a characteristic feature of a finite-size reflecting surface aperture radiating in the far-field, a reflecting surface can also be designed to achieve beam-focusing in the near-field. This can be realized by appropriately adjusting the reflection phases of the unit cells making the reflecting surface concentrate its radiation over a certain field of view (FoV) in a 3D space. This feature of a reflecting surface goes beyond the beam-steering capability in that by focusing the radiated field over a constrained FoV, we can not only control the direction of the radiated wavefront (as in beam-steering) but also the depth of the maximum radiation in the direction of propagation (along the depth). Focusing electromagnetic waves over a certain focal spot brings unique advantages, such as increased efficiency for wireless power transfer applications, higher sensitivity in non-destructive evaluation and Radio Frequency Identification (RFID), and superior channel characteristics for high data rate wireless communications to name a few \cite{7912361, buffi2010focused, stephan2007near, smith2017analysis, yu2018design}. 
Focusing the aperture radiated wavefront over a focal point can be achieved within the radiative near-field region of a reflecting surface, which is also known as the Fresnel region. The distance for the radiative near-field region of a finite size aperture with a size $D$ falls within the following interval \cite{smith2017analysis, gowda2016wireless}: 

\begin{equation}
    0.62\sqrt{D^3/\lambda_c}<d<2D^2/\lambda_c, 
    \label{Equation_5}
\end{equation}
where $\lambda_c$ is the wavelength. In (\ref{Equation_5}), the lower-bound limit, $0.62\sqrt{D^3/\lambda_c}$, is known as the boundary at which the aperture radiated fields transition from the reactive near-field region to the radiative near-field region. Beyond this region, evanescent fields are not present. The upper-bound limit in (\ref{Equation_5}), $d<2D^2/\lambda_c$ is known as the far-field (or Fraunhofer) region. In the far-field, ideally at an infinite distance, it can be assumed that the reflected wavefronts from the individual unit cells forming the reflecting surface exhibit parallel rays. Therefore, unlike beam-steering, the beam-focusing concept can only be achieved in the radiative near-field region. The design of the focus begins with defining a focal point in the radiative near-field region of the reflecting surface. Treating this point as a virtual point source located at $r''(x'',y'',z'')$ and defining $r''(x'',y'',z'')$ as the origin, the back-propagated field on the aperture plane, $r(x,y,z)$, can be calculated as follows:

\begin{equation}
    E_{\textrm{Aperture}}(r)=\frac{e^{jk_0|r-r''|}}{4\pi|r-r''|}.
    \label{Equation_6}
\end{equation}

Using the same backscatter patch as the reference-wave source calculated in (\ref{Equation_1}), $E(r)$, the required phase distribution on the reflecting surface can be achieved by interacting the projected virtual point source on the aperture plane in (\ref{Equation_6}), $E_{\textrm{Aperture}}(r)$, and the reference-wave in (\ref{Equation_2}), $E(r)$, as follows: 

\begin{equation}
    \xi_{\textrm{unit}}(x_i,y_i)=\angle{E(x_i,y_i)E^*_{\textrm{Aperture}}(x_i,y_i)}.
    \label{Equation_7}
\end{equation}

The produced interference pattern is a phase hologram, representing the phase distribution of the unit cells forming the reflecting surface. Using the holographic principle described earlier, interacting the incident reference-wave with the calculated phase pattern ensures that the reflected wavefront focuses at the focal point, which is originated from the virtual point source. As an example, in Fig. \ref{Figure_2}, we demonstrate a reflecting surface focusing at $d=0.45$m distance in the broadside direction ($\theta=0$, $\phi=0$) of the reflecting surface aperture. Taking a cross-section of the range (or depth) profile along the $z=0$ axis, the increased field intensity at the focal point of the aperture is evident.

\section{Spatial Modulation with a Near-Field Focusing IRS}

\begin{figure}[h!]
	\centering
	\includegraphics[width=6.5 cm,trim=0cm 0cm 0cm 0cm]{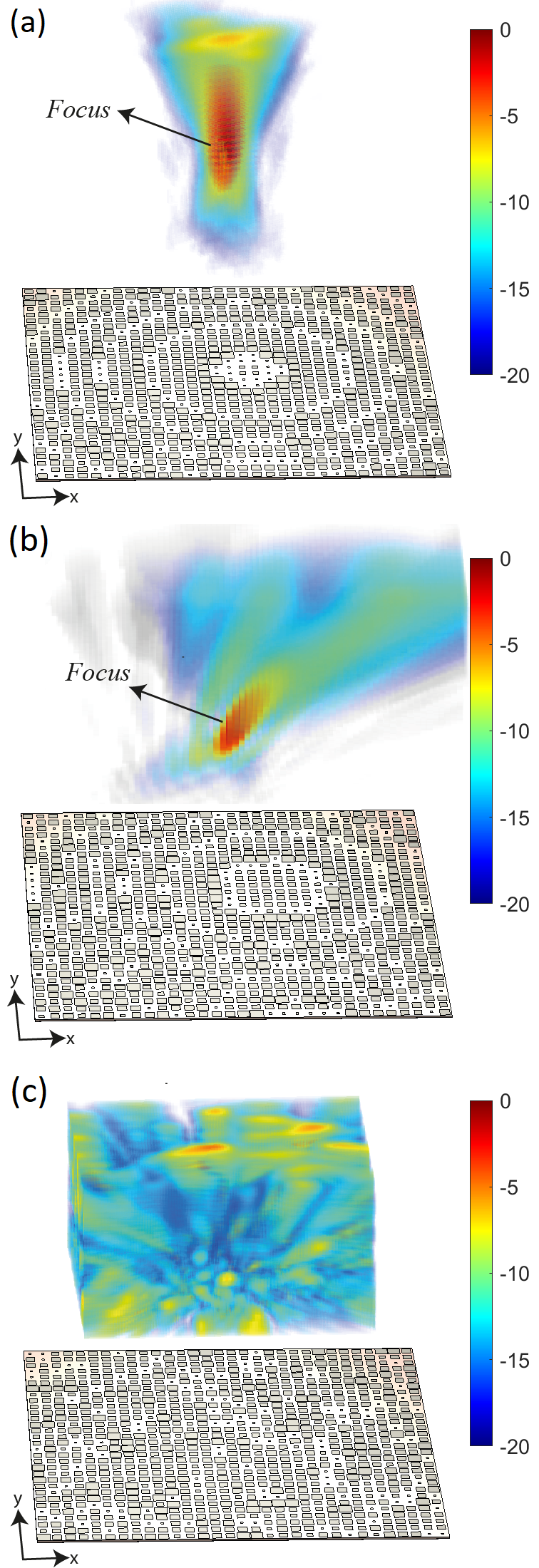}
	\caption{The radiated E-field patterns for the reflecting surface with the unit cell phase responses assigned (a) to focus at the receiver in ($\theta_1=0^\circ$, $\phi_1=0^\circ$) direction at $d_1=0.6$m focal distance (b) to focus at the receiver in ($\theta_2=15^\circ$, $\phi_2=15^\circ$) direction at $d_2=0.4$m focal distance and (c) defocused (randomized phase distribution for the unit cells). Colorbar is in dB scale.}
	\label{Figure_4}
\end{figure}

In a backscatter environment using an available secondary source as the reference-wave impinging on the reflecting surface, we can leverage a dynamic modulation principle to reconfigure the phase response of the unit cells across the reflecting surface to focus the electromagnetic radiation to an arbitrarily selected focal point in 3D space. This dynamic reconfigurablity of a reflecting surface, which produces the intelligent reflection concept and has conventionally been studied as a far-field beam-steering problem \cite{Wu,bjornson2019intelligent,8723525,9087848,8742603}, is particularly important to establish a continuous communication link with the receiver unit which can be placed at an arbitrary location within the FoV of the reflecting aperture. The near-field focusing technique can be used as an enabling technology to achieve a spatial modulation scheme over a receiver unit placed at the focal point of the reflecting surface. As an enabling technology to dynamically reconfigure the reflection phase of the aperture, the unit cells can be loaded with semiconductor elements \cite{7939976,7448838,clemente2013wideband}. As an interesting application example, in Figs. \ref{Figure_4}a and \ref{Figure_4}b, we place a receiver unit at the focal point of the reflecting surface at arbitrarily selected locations, ($\theta_1=0^\circ$, $\phi_1=0^\circ$) at a focal distance $d_1=0.6$m, and ($\theta_2=15^\circ$, $\phi_2=15^\circ$) at a focal distance $d_2=0.4$m. In Fig. \ref{Figure_4}c, we study a chaotic case, which does not produce a focus within the FoV of the reflecting surface. For these three cases, we use the same backscatter patch antenna as the source of the reference-wave, and define a threshold, -10 dB on the normalized E-field magnitude at the receiver. In this study, we use a single reflecting surface aperture to dynamically focus and defocus the radiated wavefront. The scenario of reading an E-field magnitude of above -10 dB threshold at the receiver constitutes to a "1" whereas a signal level below the selected threshold at the receiver, constitutes to a "0". While the focusing can be achieved by choosing the correct reflection phases for the unit cells as explained in Section II.B, the defocusing of the radiated field can be achieved by randomizing the unit cell phase responses across the reflecting surface aperture. The 3D E-field patterns for the reflecting surface with focused and randomized unit cell phase responses are shown in Fig. \ref{Figure_4}.

\begin{figure}[t!]
\centering
\includegraphics[width=8.5 cm,trim=0cm 0cm 0cm 0cm]{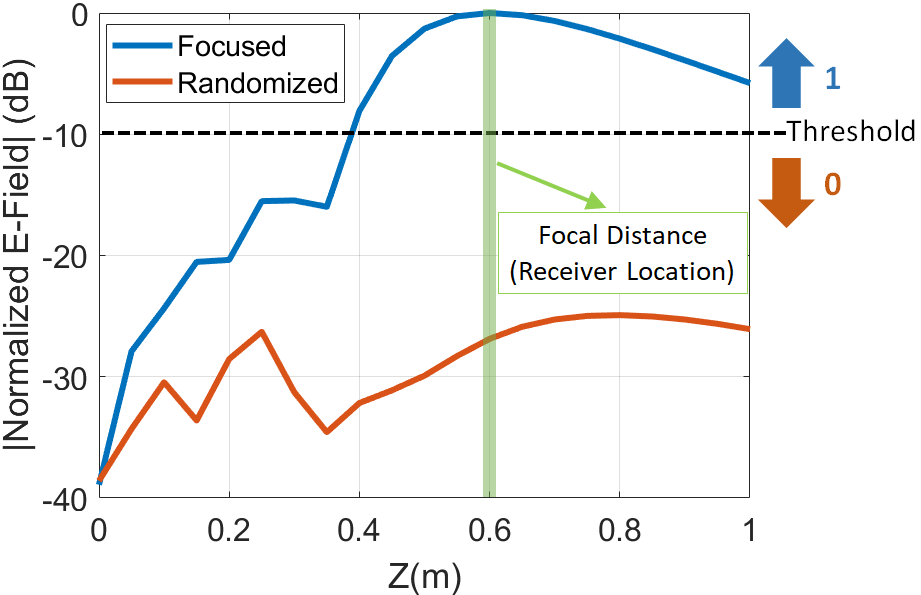}
\caption{Spatial modulation principle at the receiver as a function of the E-field strength. Focusing the E-field results in the signal at the receiver exceeding the selected threshold producing "1". Randomizing the unit cell phase responses results in a chaotic wavefront resulting in the signal at the receiver remaining below the selected threshold level producing "0".}
\label{Figure_5}
\end{figure}

As can be seen in Fig. \ref{Figure_4}a, by adjusting the reflection phase response of the unit cells across the reflecting surface to achieve a constructive interference of the radiated wavefront in the direction ($\theta_1=0^\circ$, $\phi_1=0^\circ$) at $d_1=0.6$m distance, the formed focal spot is evident. Similarly, moving the focal point to a distance of $d_2=0.4$m in the direction ($\theta_2=15^\circ$, $\phi_2=15^\circ$), the reflecting surface can reconfigure the reflection phase response across the unit cells to shift the focal point accordingly as shown in Fig. \ref{Figure_4}b. In comparison to these constructive interference cases, by randomizing the phase responses of the unit cells across the reflecting surface, we create a chaotic wavefront as shown in Fig. \ref{Figure_4}c, which is devoid of a well-defined focus, reducing the E-field magnitude at the receiver significantly. To provide a comparison between the E-field magnitude patterns along the trajectory of the receiver, we consider the broadside ($\theta_1=0^\circ$, $\phi_1=0^\circ$) focusing scenario at $d_1=0.6$m distance in Fig. \ref{Figure_4}a and the defocusing scenario with a randomized phase distribution in Fig. \ref{Figure_4}c as a reference. The E-field magnitude patterns for these two cases along the depth axis of the reflecting surface (z-axis) is shown in Fig. \ref{Figure_5}.

Analyzing Fig. \ref{Figure_5}, it can be seen that by focusing the radiation of the IRS at the receiver, the E-field magnitude is increased by 27 dB in comparison to the defocused IRS with randomized unit cell phase responses. From Fig. \ref{Figure_5}, it is evident that we can distinguish between the "1" and "0" cases by simply focusing and defocusing the wavefront distribution of the reflecting surface illuminated by a backscatter tag as the reference-wave. It is important to emphasize that using an IRS, this dynamic modulation principle can be realized in an all-electronic manner, and the focusing of the radiated electromagnetic wavefront can be achieved at an arbitrary focal distance, $d$, in an arbitrary direction ($\theta, \phi$) within the FoV of the reflecting surface. This principle relies entirely on the holographic beam forming principle, does not require mechanical scanning of the FoV, and eliminates the need for hardware-intense phase shifting circuits to achieve the dynamic modulation of the radiated wavefront, differentiating this principle from the well-known phased array technology \cite{mailloux2017phased}. 

In Sections II and III, the radiated wavefront of the studied reflecting surfaces is modified by means of adjusting the reflection phase response of the unit cells forming the reflective aperture. As an interesting alternative, a time delay-based approach can also be adopted to modulate the wavefront of an IRS aperture. This is presented in the next section.     

\section{Backscattering through TIme-Varying IRS}
\begin{figure}[t!]
	\centering
	\subfloat[]{
		\includegraphics[width=0.45\linewidth,valign=m]{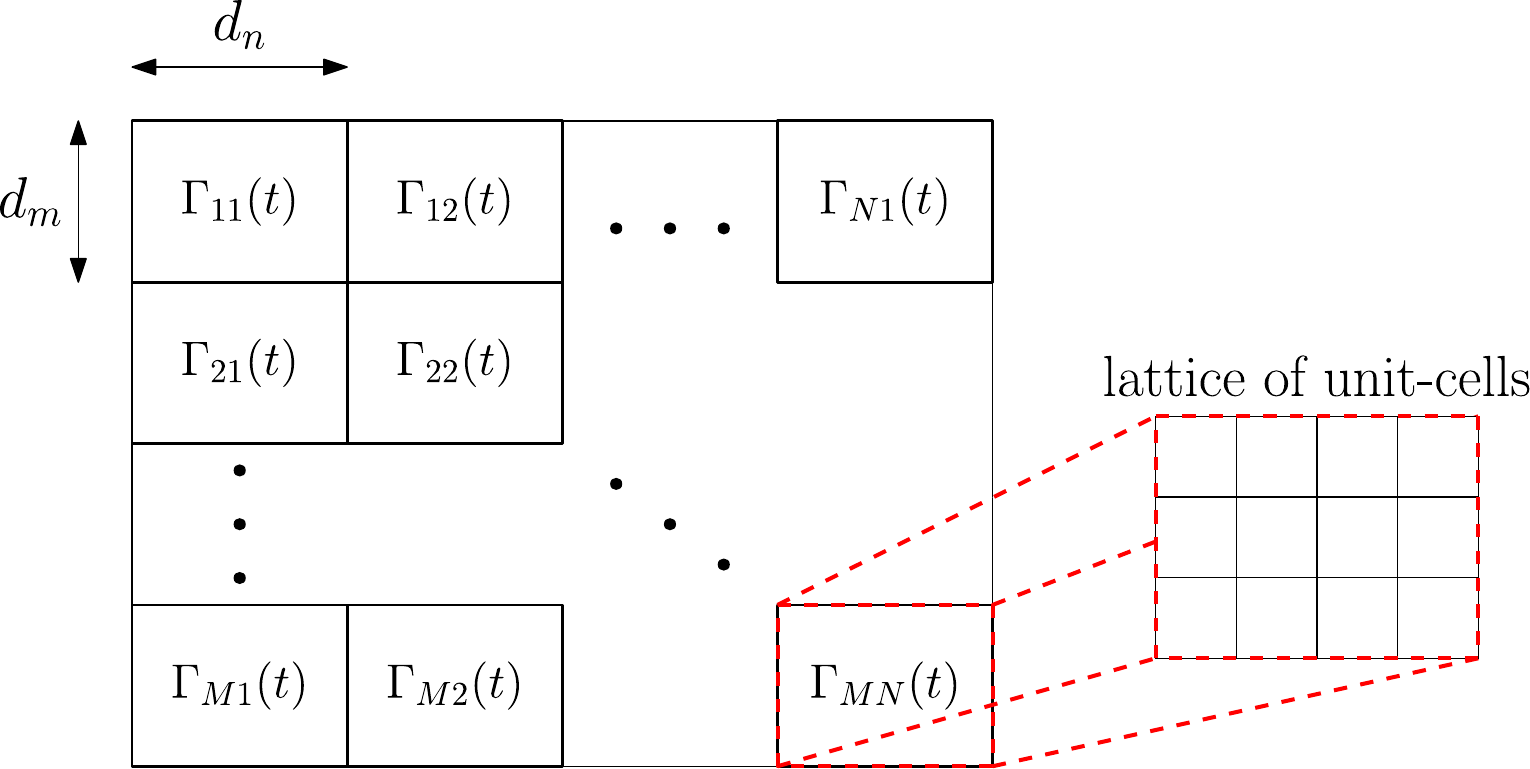}
	\label{fig:fig001A}
	}
\hfil
	\subfloat[]{
	\includegraphics[width=0.45\linewidth,valign=m]{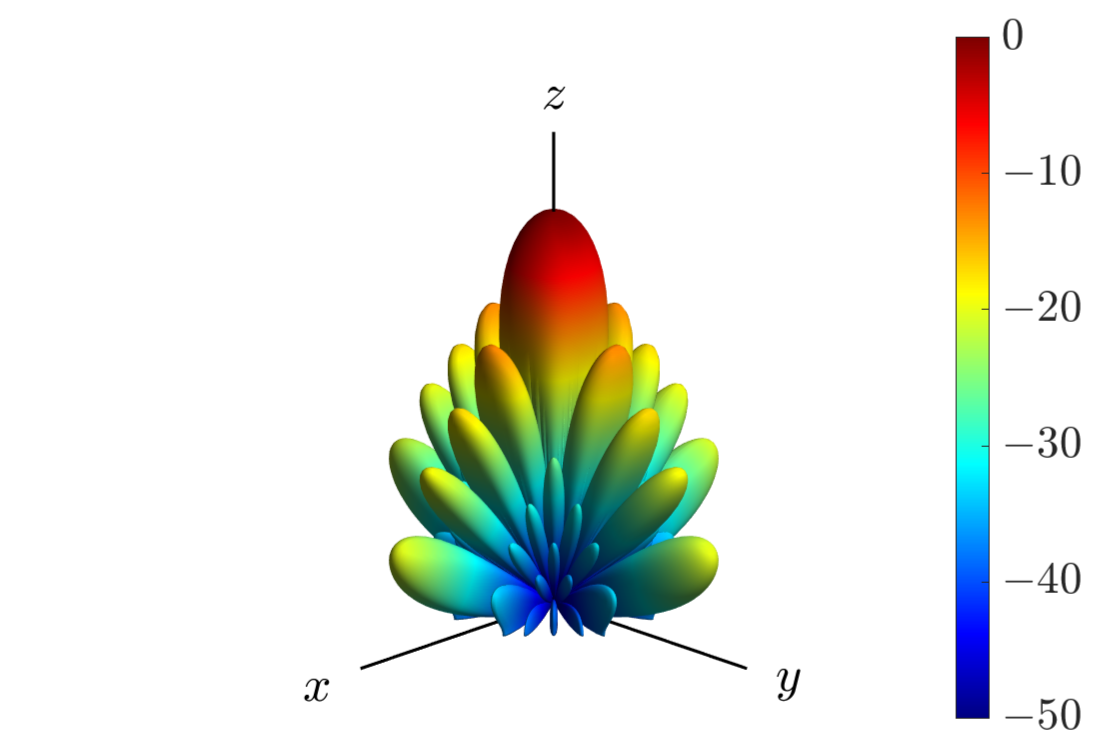}
	\label{fig:fig001B}
	}
	\caption{(a) A typical representation of a planar, time-varying IRS consisting of $M\times{}N$ lattices, each of those is represented by the time varying reflection coefficient $\Gamma_{mn}(t)$). (b) For time constant $\Gamma_{mn}=1$, the backscattered electrical field $E(\theta,\phi)$ in dB scaling is also depicted.}
	\label{fig:fig000}
\end{figure}

A geometric representation of a planar, time varying IRS is depicted in Fig. \ref{fig:fig001A}:
it  consists of $ M \times N $ orthogonal lattices of dimension $ d_m \times d_n $.
Each lattice corresponds to a time varying reflection coefficient $ \Gamma_{mn}\left( t \right)  $.
It is noted, that each lattice could be formed of a periodic structure of unit cells with constant reflective coefficient \cite{Su2016,Assimonis2019}.
For normal incidence of time harmonic plane waves of frequency $ f_c $, the normalized scattered by the RS electrical field in the far-field region is given by \cite{Zhang2018},
\begin{equation}\label{antenna_array_radiation_pattern}
\begin{split}
	E \left(t, \theta, \phi \right) = 
	\frac{E_0 \left( \theta, \phi \right)}{M\,N}
	\sum_{m=1}^{M}  \sum_{n=1}^{N} 
	{\Gamma}_{mn}\left( t \right)  \,
	F_{mn}\left(\theta,\phi \right),
\end{split}
\end{equation}
with
\begin{equation}
    F_{mn}\left(\theta,\phi \right) = e^{ j \frac{2\pi }{c_0} f_c
		\left\lbrace  
					\left(m-1\right) d_m \sin(\theta)\cos(\phi) +  \left(n-1\right) d_n \sin(\theta)\sin(\phi)  
		\right\rbrace  
	},
\end{equation}
where $ c_0 $ denotes the propagation speed,
$F_{mn}\left(\theta,\phi \right)$ is equivalent to $AF\left(\theta,\phi \right)$ in \eqref{Equation_2} and
$ E_0 \left( \theta, \phi \right) $ is the far-field radiation pattern of each lattice. Assuming isotropic elements, it is
	$ E_0 \left( \theta, \phi \right) = 1 $
%
and for $ M=N=10 $ IRS with $ d_m = d_n = \lambda_c/2 $ and invariant reflection coefficients (i.e., $ \Gamma_{mn} = 1 $) the $ E \left(\theta, \phi \right) $ is depicted in Fig. \ref{fig:fig001B}: the maximum field occurs above the RS at position $ \theta = 0^\circ$, $ \phi = 0^\circ$. 

In a typical time-varying, binary backscatter system, reflection coefficient $ \Gamma\left(t \right)\ $ periodically alternates between two states ($ \Gamma^{(1)}, \Gamma^{(2)} $),  modulating information \cite{Assimonis2016}. 
In order to minimize the  error probability  $ P_e $ of such a  system,  the Euclidean distance between  $ \Gamma^{(1)}, \Gamma^{(2)} $ should be maximized \cite{Bletsas2010,Assimonis2014}, therefore,
\begin{equation}
	\!\min_{\Gamma^{(1)},\Gamma^{(2)}}\left\lbrace P_e \right\rbrace
	\Leftrightarrow
	\!\max_{} \left\lbrace \left| \Gamma^{(1)} - \Gamma^{(2)} \right|  \right\rbrace.
\end{equation}
Two possible solutions in the unit-circle (Smith chart) are 
\begin{equation}
\begin{aligned}
	\Gamma^{(1)} &= \phantom{+}1 \quad \mathrm{(open-circuited)}\\
	\Gamma^{(2)} &= -1					\quad \mathrm{(short-circuited)}\
\end{aligned}
\end{equation}
where, $ \left| \Gamma^{(1)} - \Gamma^{(2)} \right| = 2 $.
Thus, in this work we assume that the reflection coefficient $ \Gamma_{mn}\left( t \right)  $ of each lattice  periodically alternates between $ \pm1 $ with frequency $ f_{0} $, hence within a period $ T_0  = 1/f_0$ it is, 
\begin{equation}\label{gamma_t}
	\gamma_{}\left( t\right) = 
	\bigg\{ 
	\left[ 
	u\left( t \right) - u\left( t -\frac{T_0}{2} \right)
	\right]
	-
	\left[ 
	u\left( t - \frac{T_0}{2} \right) - u\left( t -T_0 \right)
	\right] 
	\bigg\},
\end{equation}
where, $ u\left( t \right)$, is \textit{Heaviside} step function \cite{NguyenShwedyk2009}, and thus,
\begin{equation}\label{Gamma_t}
\Gamma_{mn}\left( t\right) = \sum_{k=-\infty}^{\infty} \gamma_{}\left( t - k T_0 - \tau_{mn}\right),
\end{equation}
where, $ k=0,\pm1,\pm2,... $ and $ \tau_{mn} $ is the time delay for each lattice: for synchronized lattices $ \tau_{mn}  = 0 $ s. Also, it should be noted that \eqref{antenna_array_radiation_pattern} is valid for $ f_c \gg f_0 $ \cite{Zhang2018}.
The \textit{Fourier expansion} of \eqref{Gamma_t} is given by \cite{NguyenShwedyk2009}
\begin{equation}\label{Fexp}
	\Gamma_{mn}\left( t\right) = \sum_{k=-\infty}^{\infty} D^k_{mn}e^{j 2 \pi k f_0 t},
\end{equation}
where,
\begin{equation}\label{Dk}
\arraycolsep=1.5pt\def\arraystretch{1.5}
	D_{mn}^{k} 
	\triangleq 
	\frac{1}{T_0} \int_{t\in T_0}^{} \Gamma_{mn}\left( t\right) e^{ - j 2 \pi k f_0 t } \, {d}t
	=
	\left\{
	\begin{array}{ll}
	\dfrac{2}{\pi k}{ e^{ - j \left( 2 \pi k f_0 \tau_{mn} + \pi/2\right)   } }    		& \quad k~\mathrm{odd} \\
							0 	 & \quad k=0,\mathrm{even}.  \\
	\end{array} 
	\right. 
\end{equation}

\begin{figure*}[t!]
	\centering
	\subfloat[]{
		\includegraphics[width=0.27\linewidth]{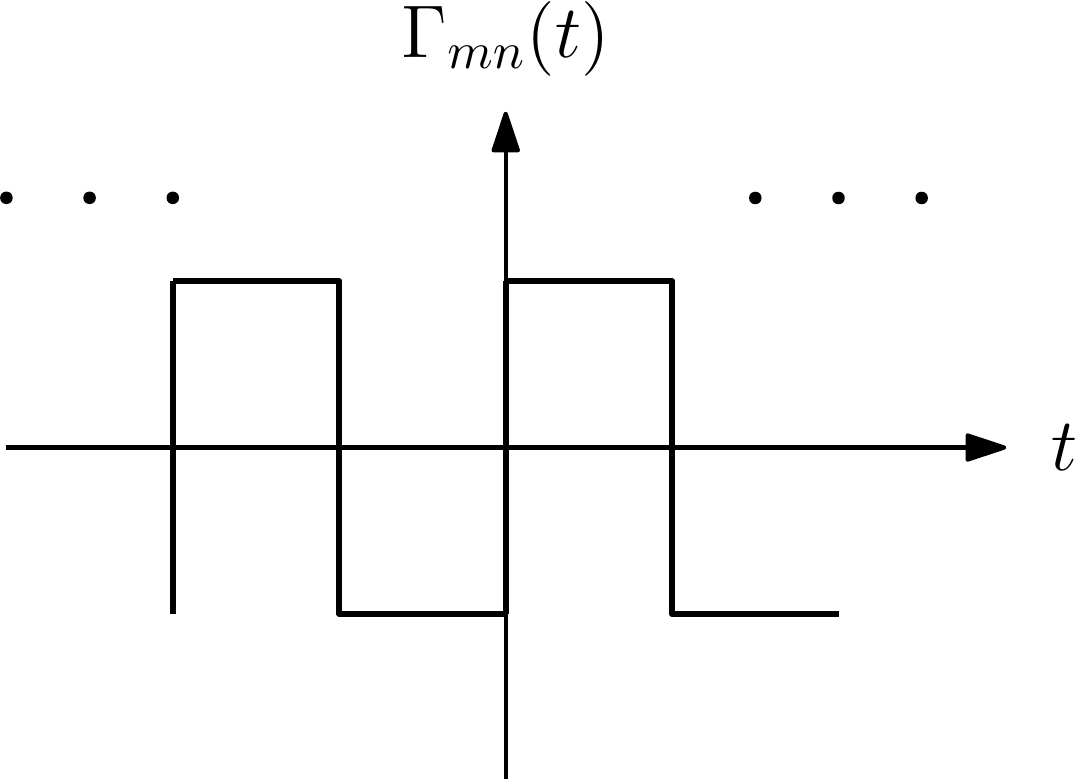} 
		\label{fig:fig002A}
	}
	\subfloat[]{
		\includegraphics[width=0.27\linewidth]{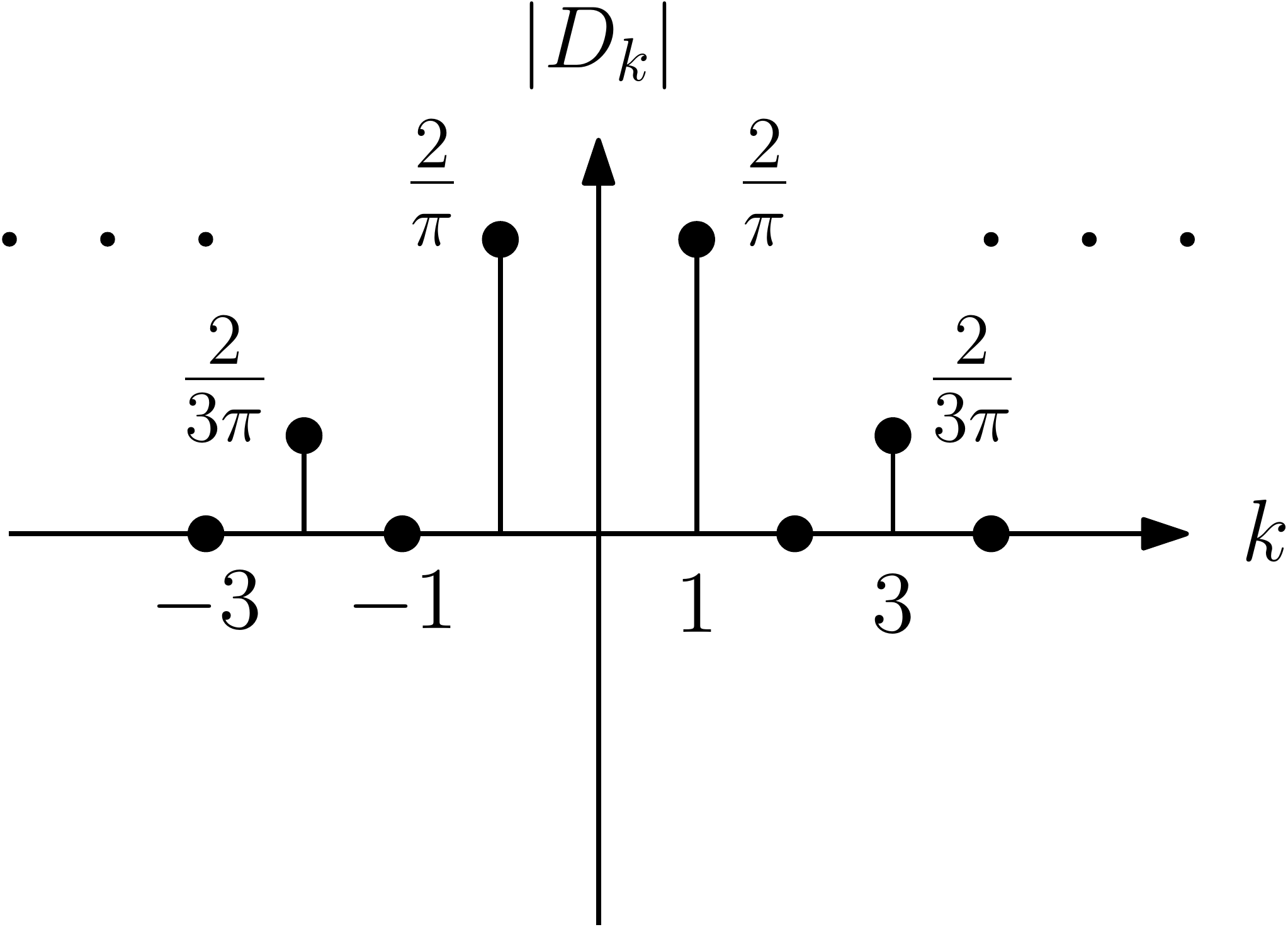}  
		\label{fig:fig002B}
	}
	\subfloat[]{
		\includegraphics[width=0.27\linewidth]{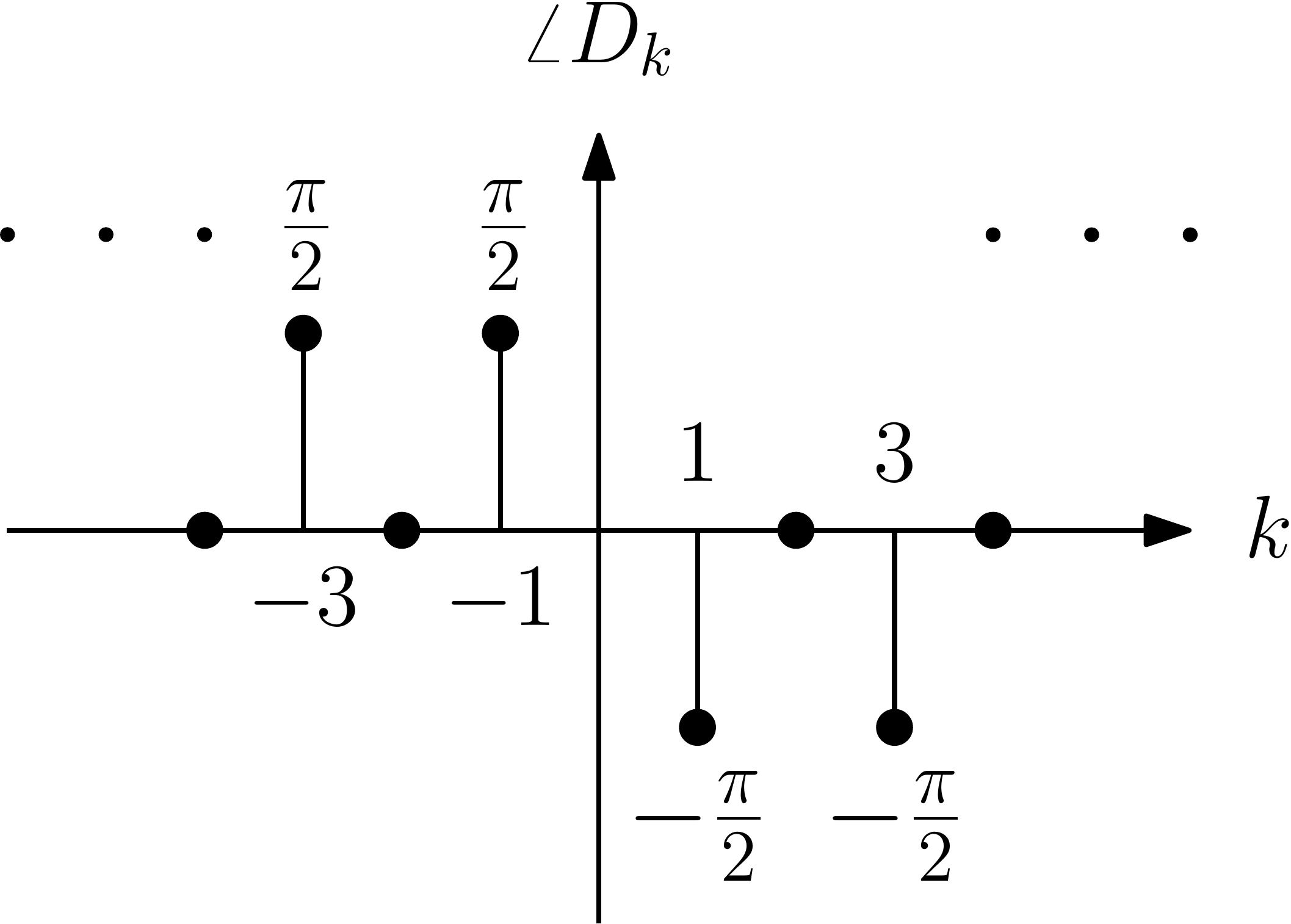}  
		\label{fig:fig002C}
	}
	\caption{
		(a) Periodic signal (square wave) of the reflection coefficient $\Gamma_{mn}(t)$ assuming zero delay between the lattices, i.e., $\tau_{mn}=0$s.
		The complex coefficient $D^k$ of the $\Gamma_{mn}(t)$ Fourier expansion: magnitude (b) and phase (c).
	}\label{fig:fig002}
\end{figure*}

Fig. \ref{fig:fig002} depicts the time varying reflection coefficients $ \Gamma_{mn}\left( t \right)  $ for $ \tau_{mn} = 0 $s and the complex coefficients $ D^k $ (magnitude and phase): the incident signal at $ f_c $ is reflected over the harmonics $ f_c \pm f_0 $. Also, the magnitude of $ D_k $'s is reduced by a factor of $ 2/k\pi $, while the phase remains constant ($ \pm \pi/2 $ for $ k $ negative and positive, respectively).

The frequency spectrum of the normalized scattered electromagnetic field is estimated through the \textit{Fourier transform} of \eqref{antenna_array_radiation_pattern}, and thus,
\begin{equation}
	\mathcal{E}^k
	= 
	\mathcal{F} 
	\left\lbrace 
	E\left(t, \theta, \phi \right)  
	\right\rbrace
	= \mathcal{E} \left(f, \theta, \phi \right)
	=
	\frac{E_0 \left( \theta, \phi \right)}{M\,N}
	\sum_{k=-\infty}^{\infty} 
	\left\lbrace
	\sum_{m=1}^{M}  
	\sum_{n=1}^{N} 
	D^k_{mn} 
	\,
	F_{mn}\left(\theta,\phi\right)
\right\rbrace
\delta\left( f - k f_0 \right),
\end{equation}
where, $ \mathcal{F}  \left\lbrace  \cdot  \right\rbrace $, denotes the Fourier transform. It is evident that, the scattered electrical field is a sequence of impulses located at the harmonics $ k f_0 $.

\begin{figure*}[t!]
	\centering
	\subfloat[]{
		\includegraphics[width=0.33\linewidth]{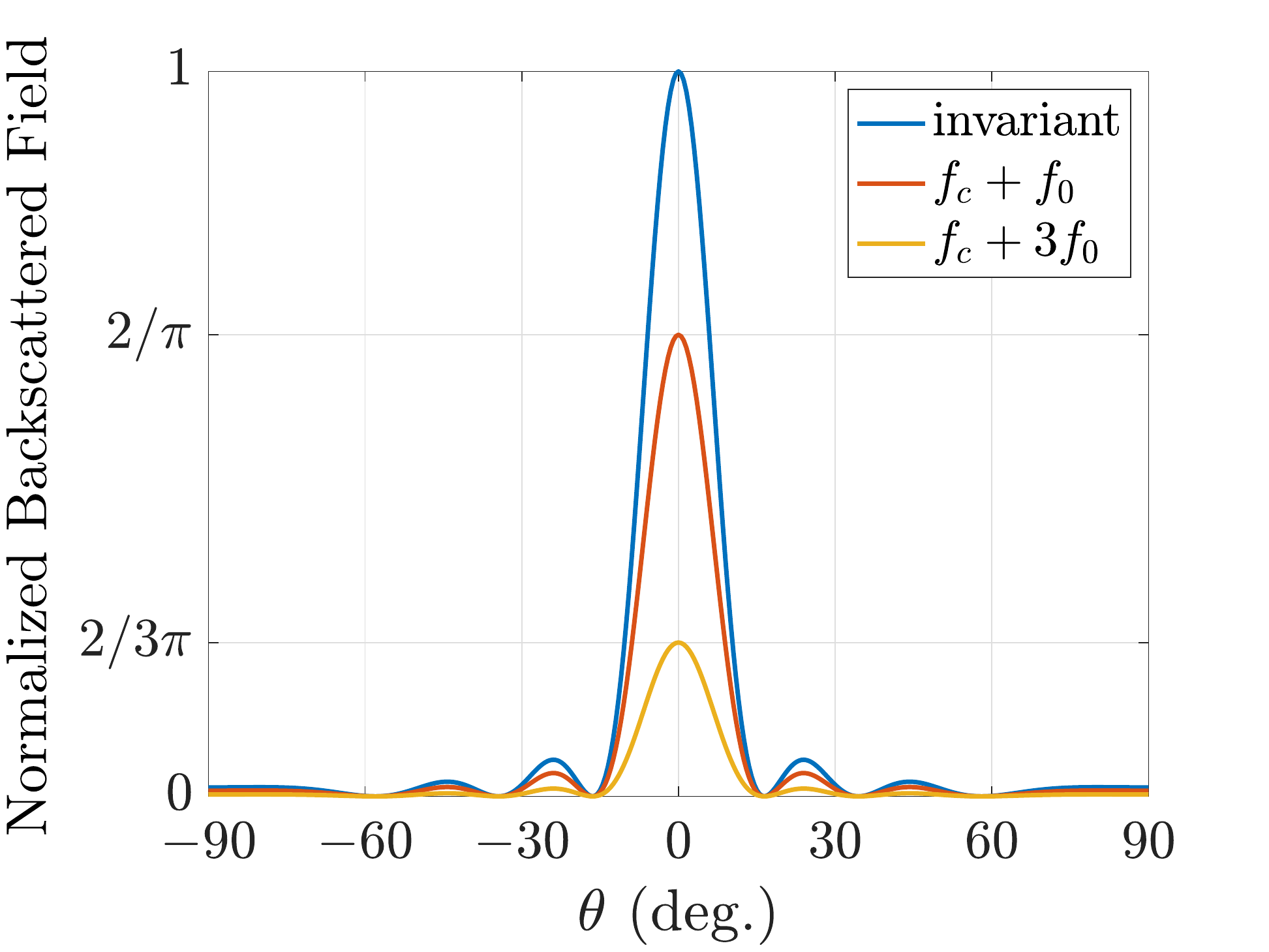}  
		\label{fig:fig003A}
	}
	\subfloat[]{
		\includegraphics[width=0.33\linewidth]{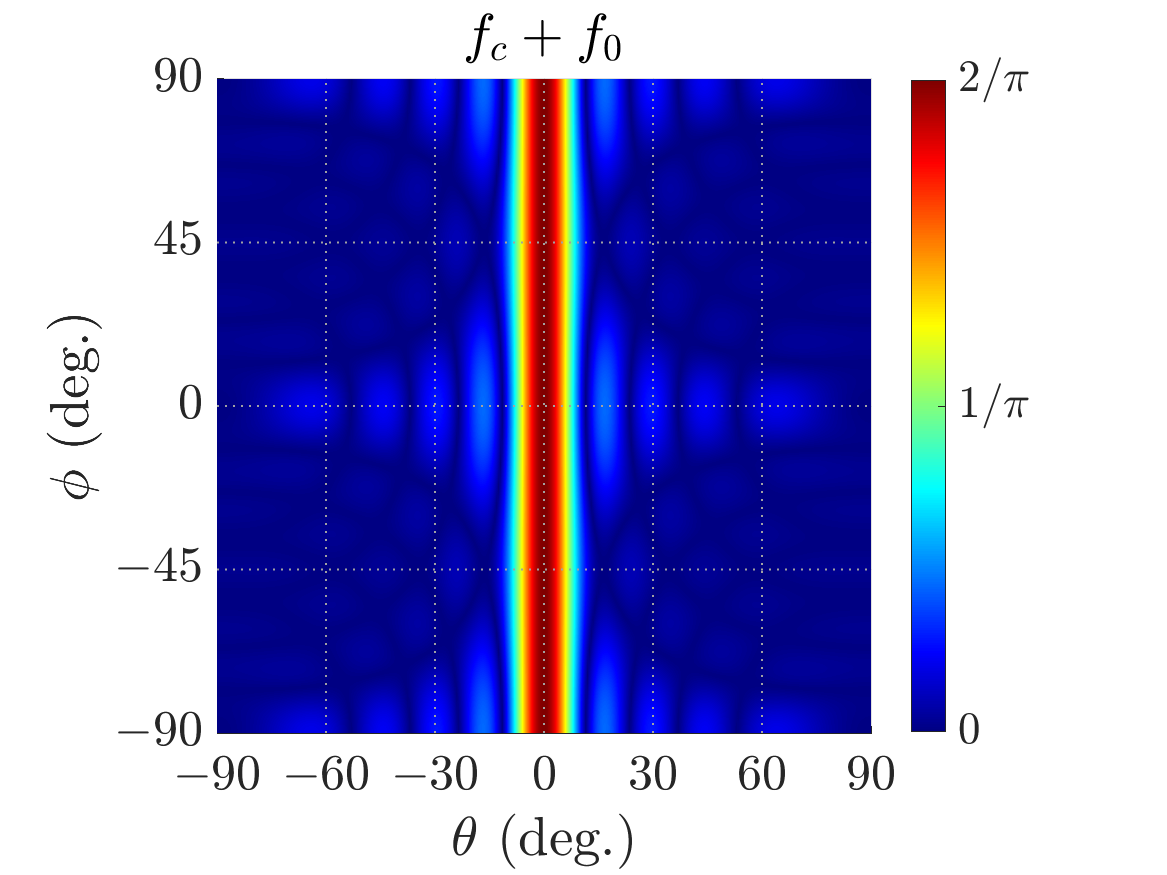}  
		\label{fig:fig003B}
	}
	\subfloat[]{
		\includegraphics[width=0.25\linewidth]{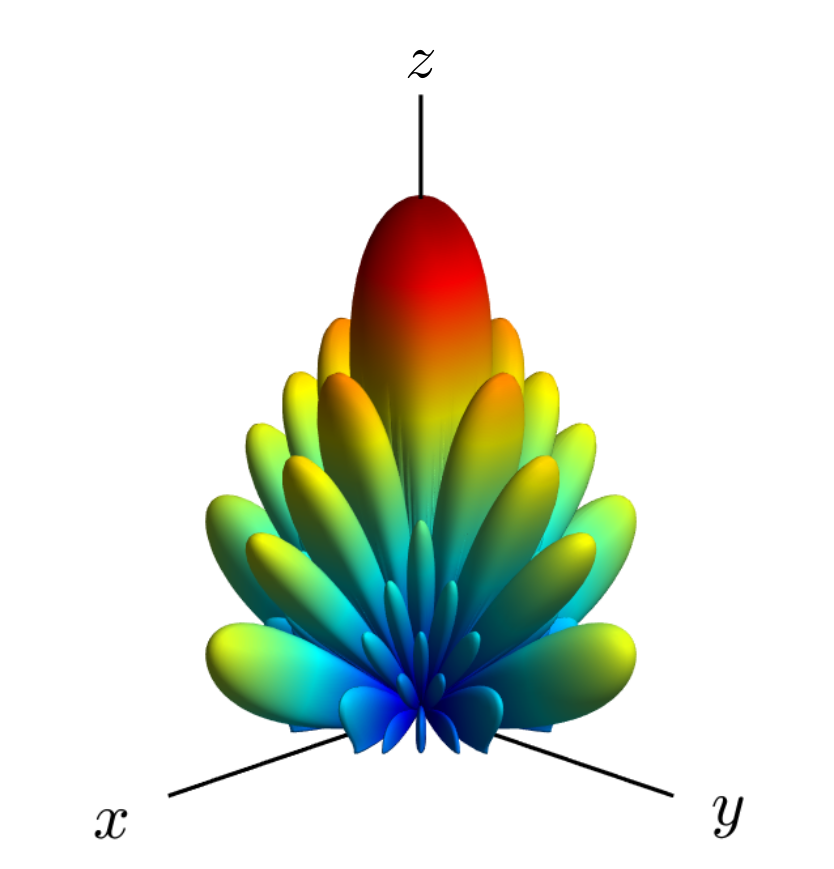}  
		\label{fig:fig003C}
	}
	\caption{
		(a) The normalized backscattered electrical field  when $\Gamma_{mn}=1$ (invariant case) and when $\Gamma_{mn}(t)$ is a square wave with $\tau_{mn}=0$s: the first ($f_c+f_c$) and second $f_c+3f_c$ harmonic is presented. 
		$\mathcal{E}^k$ is also depicted (only the first harmonic) versus $\theta,\phi$, in linear (b) and dB scaling (c).
		Since $\tau_{mn}=0$, energy is reflected back to the  direction of the incident wave.
	}
	\label{fig:fig003}
\end{figure*}
\begin{figure*}[!t!]
	\centering
	\subfloat[]{
		\includegraphics[width=0.33\linewidth]{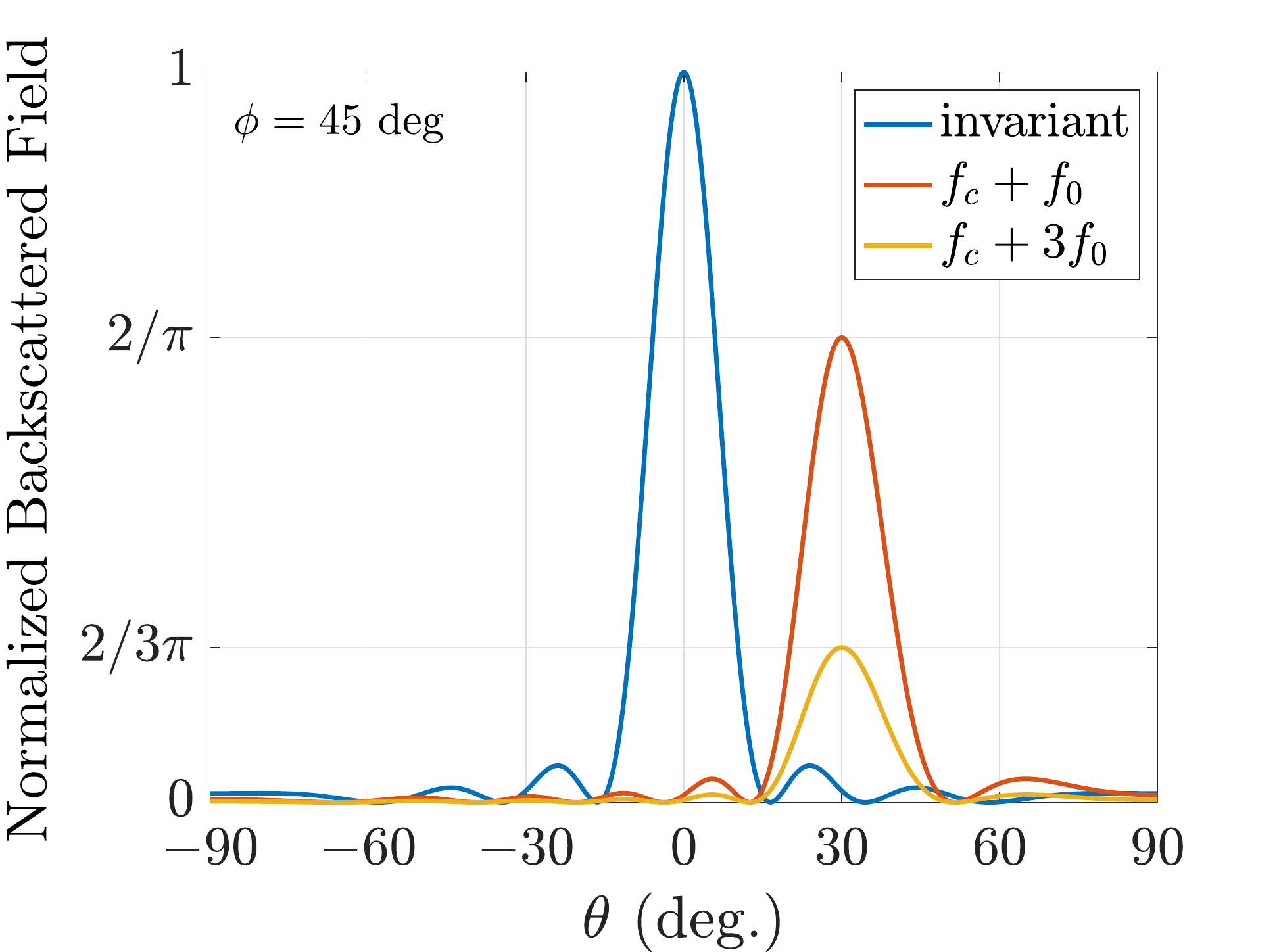}  
		\label{fig:fig004A}
	}
	\subfloat[]{
		\includegraphics[width=0.33\linewidth]{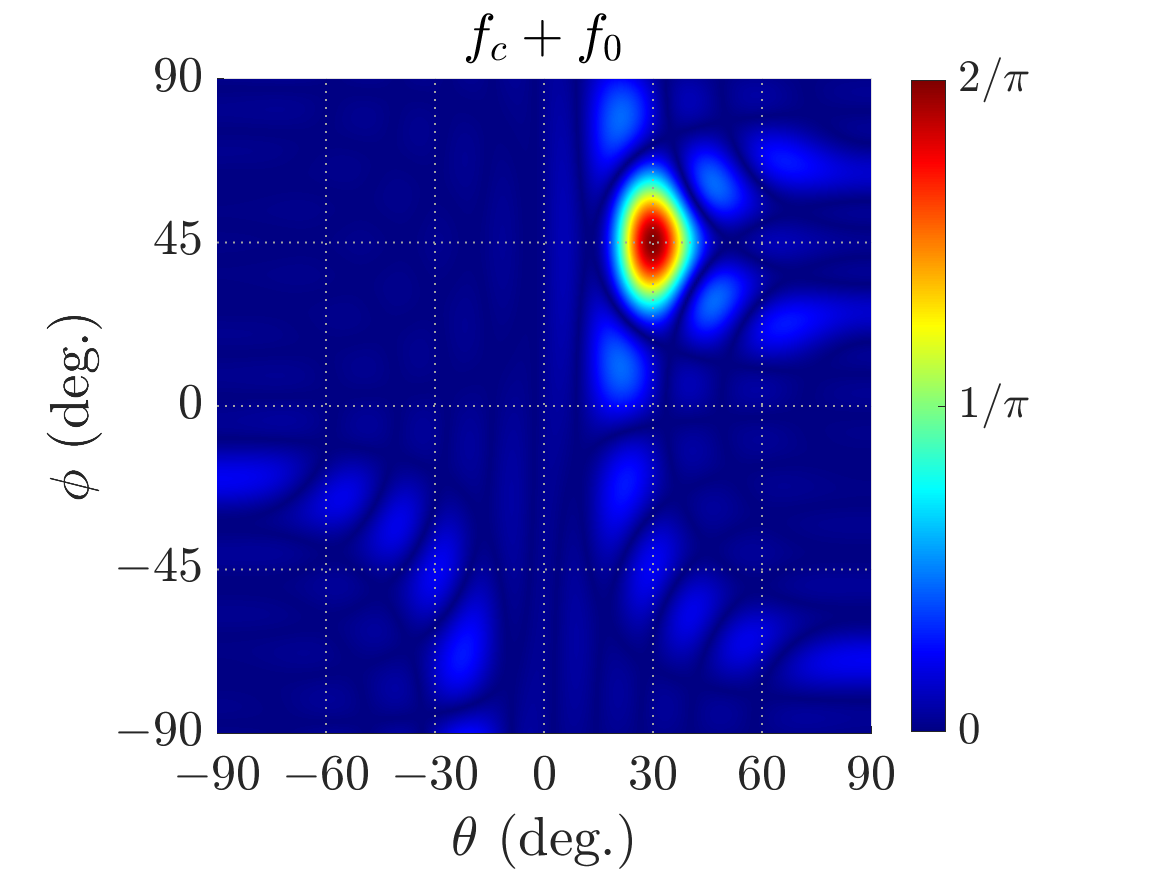}  
		\label{fig:fig004B}
	}
	\subfloat[]{
		\includegraphics[width=0.25\linewidth]{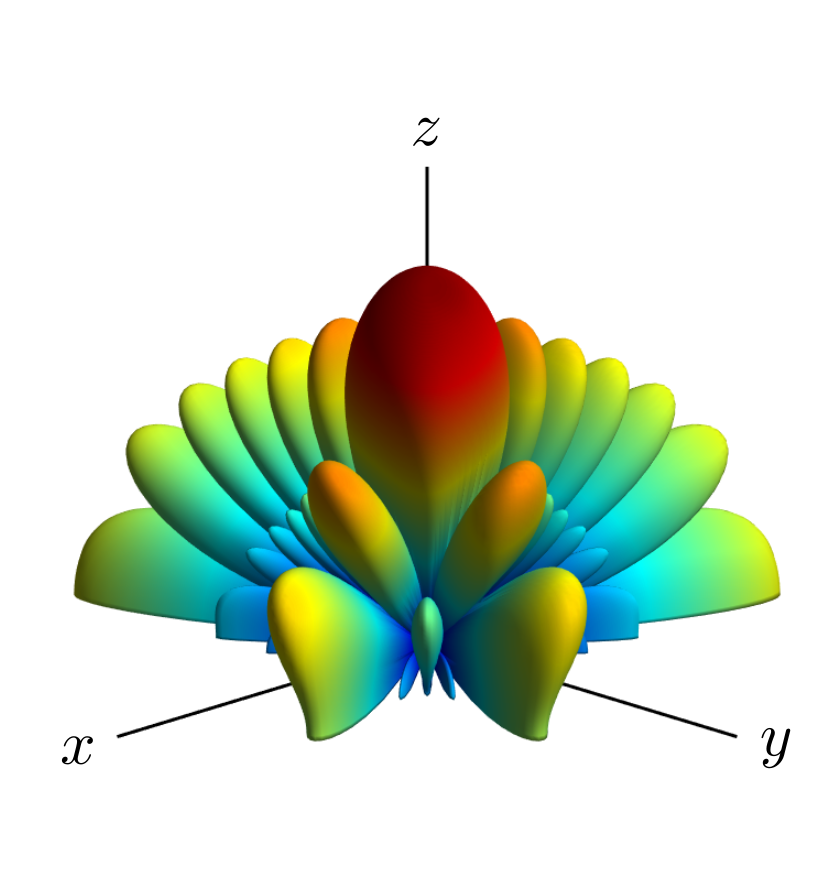}  
		\label{fig:fig004C}
	}
	\caption{
		(a) The normalized backscattered electrical field  when $\Gamma_{mn}=1$ (invariant case) and when $\Gamma_{mn}(t)$ is a square wave with $\tau_{mn}$ estimated through \eqref{tau_focus} for $\theta_0=30^\circ$ and $\phi_0=45^\circ$: the first ($f_c+f_c$) and second $f_c+3f_c$ harmonic is presented. 
		$\mathcal{E}^k$ is also depicted (only the first harmonic) versus $\theta,\phi$, in linear (b) and dB scaling (c).
		Energy is reflected back to the desired direction, i.e., $(\theta_0,\phi_0)$.
	}
	\label{fig:fig004}
\end{figure*}

Fig. \ref{fig:fig003A} depicts the $ E  $ ($ \Gamma_{mn} = 1$, i.e., invariant case) and the $ \mathcal{E}^k $ over $ \theta $ (it is $ \phi = 0^\circ$). for the first ($ k=1 $) and second ($ k=3 $) harmonic  and for $ \tau_{mn} = 0$s (i.e., for all lattices, $ \Gamma_{mn}(t) $ periodically alternates between $ \pm 1 $ at the same time):  the incident signal is reflected back to the same direction, however with reduced magnitude. The latter, results from the behaviour of the complex coefficients $ D^k $, which are $ \angle D^k = -\pi/2 $ (constant), however $  |D^k| = 2/k\pi$ for $ k=1,3 $, as mentioned above. Fig.  \ref{fig:fig003B}, \ref{fig:fig003C} also depict the $ \mathcal{E}^k $ over $ \theta, \phi $ for only the first harmonic. 

The delay coefficients  $ \tau_{mn} $ can be used for beamforming, steering the beam to specific direction $ (\theta_0,\phi_0) $ by applying,
\begin{equation}\label{tau_focus}
	\tau_{mn} =
	 \frac{f_c}{c_0 \, k f_0} 
\bigg\{
			 \left(m-1\right) d_m \sin(\theta_0)\cos(\phi_0) 
			 +
			 \left(n-1\right) d_n \sin(\theta_0)\sin(\phi_0)
\bigg\}.
\end{equation}

For $ \theta_0=30^{\circ}$, $\phi_0=45^{\circ} $ the  $ \mathcal{E}^k $ is estimated  through \eqref{Dk}, \eqref{tau_focus}  for the first and the second harmonic and the results are depicted in Fig. \ref{fig:fig004A}. Again, the $\mathcal{E}^k $ is reduced by a factor of $ 2/k\pi $ over the harmonics, as expected. However, now the IRS steers the reflected power to the direction $ \theta_0=30^{\circ}, \phi_0=45^{\circ} $. 
The latter is also illustrated in Fig. \ref{fig:fig004B}, \ref{fig:fig004C}, where the $ \mathcal{E}^k $-first harmonic is plotted over $\theta,\phi$.

To sum up, by properly inserting  time delay between the $ \Gamma_{mn}\left( t\right)  $, the resulting RS is able to steer the reflected power to a specific direction. 
Practically speaking, this could be implemented through only the programming of a microcontroller, which will control the IRS and not through the sophisticated design of the unit cells of the lattices.
In this work, unit cells are designed once with their reflection coefficients, which periodically alternate between $ -1,1 $ and then, a microcontroller caters for beamforming. This technique significantly reduces the design complexity and cost of an IRS.

\section{Conclusion}
In this paper, we have provided a detailed analysis on the design and use of holographic reflecting surfaces in a backscatter wireless communication environment. We have shown both the near-field and far-field radiation characteristics of such surfaces, diverging from the commonly adopted far-field implementation of these apertures in the literature. Specifically, we have presented that, in addition to providing a high-fidelity beam-steering performance in the far-field, the adoption of a reflecting surface in the radiative near-field brings a new degree of freedom in a communication link, which is the ability to control the strength of the aperture radiated fields at the receiver location. This goes beyond the conventional beam-steering capability of such surfaces, enabling them to be used as an active modulator in a wireless communication link. As a particular application example, we have designed a reflecting surface illuminated by a patch antenna type backscatter tag and shown that by dynamically adjusting the phases of the unit cells forming the reflecting surface, we can focus and defocus the radiated wavefront over a constrained FoV in 3D space, and its application as a spatial modulation technique. We have also developed a novel modulation technique suitable for backscattering through time-varying IRS. The main idea is to make the reflection coefficients alternate between -1 and +1 by means of a time-delay in order to generate anomalous reflections, thereby offering full flexibility and scalability to the IRS. 

Apart from backscattering applications, the presented spatial modulation techniques can also be leveraged to improve the physical layer security in a wireless communication environment by encoding certain data sequences into the transmission from the reflecting surface by means of performing a set of simple focusing/defocusing operations. The presented concepts can be achieved in an all-electronic manner eliminating the need for mechanical scanning and do not require hardware-intense, power-hungry phase shifting circuits to achieve the presented spatial modulation principle. 

\bibliographystyle{IEEEtran}
\bibliography{References.bib}

\end{document}